\documentstyle[prb,eqsecnum,aps]{revtex}

\font\bba=msbm10 scaled 1080
\font\bbb=msbm8 %scaled 1080
\font\bbc=msbm6 %scaled 1080
\newfam\bbfam
\textfont\bbfam=\bba
\scriptfont\bbfam=\bbb
\scriptscriptfont\bbfam=\bbc
\def\bb{\fam\bbfam\bba}
\def\R{{\bb R}}
\def\Z{{\bb Z}}

\begin{document}
\draft
\title
{Sine-Gordon theory for the equation of state\\ of
 classical hard-core Coulomb systems.\\
I Low fugacity expansion.}
\author{Jean-Michel Caillol \thanks{Electronic mail: 
Jean-Michel.Caillol@labomath.univ-orleans.fr}}
\address{MAPMO - CNRS (UMR 6628)\\
         D\'epartement de Math\'ematiques \\
         Universit\'e d'Orl\'eans, BP 6759\\
         45067 Orl\'eans cedex 2, France}
\author{Jean-Luc Raimbault \thanks{Electronic mail:
raimbaut@lpmi.polytechnique.fr}}
\address{CRMD - CNRS (UMR 6619) \\
Universit\'e d'Orl\'eans \\
45071 Orl\'eans cedex, France}      
         
\date{\today}
\maketitle
\begin{abstract}
We present an exact field theoretical representation of the statistical
mechanics of classical hard-core Coulomb systems.
This approach generalizes the
usual sine-Gordon theory valid for point-like charges or lattice systems to
continuous Coulomb fluids with additional short-range interactions. This
formalism is applied to derive the equation of state of the restricted primitive
model of electrolytes in the low fugacity regime up to order $\rho^{5/2}$
($\rho$ number density). We recover the results obtained by  Haga by
means of Mayer graphs expansions.
\end{abstract}

%%%%%%%%%%%%%%%%%%%%%%%%%%%%%%%%%%%%%%%%%%%%%%%%%%%%%%%%%%%%%%%%%%%%%%%%%%%%%%%%

\section{Introduction}
\label{intro}
 The aim of this paper and of the following one is to present a formally exact
field theory which allows for the calculation of thermodynamic functions of classical 
hard-core Coulomb systems. It is well known that the grand-canonical partition
function of the Coulomb gas can be represented by a sine-Gordon 
action\cite{Edwards,Kac,Siegert}. However this mapping is applicable only
for pointlike charges, or for lattice systems.\cite{Fisher0} If necessary, the short-range
 repulsion is frequently included
post-facto by introducing a suitable cutoff in momentum space integrals. In this
work, we derive a formally exact sine-Gordon field theory whitout the use of any
arbitrary cutoff. This off-lattice formalism is used in two complementary directions.
In the present paper we recover the low fugacity expansion of the thermodynamic
functions
obtained years ago by Mayer\cite{Mayer}
and Haga\cite{Haga} by means of graph resummation techniques.
 In the companion paper,
we consider rather the high temperature regime and the results obtained by Stell
and Lebowitz\cite{Stell} in the frame of the 
so-called $\gamma$-ordering theory
 are also recovered.

The simplest sound theory of electrolytes is due to Debye and H\"uckel 
\cite{Debye}
who showed, nearly eighty years ago, that, at least in the low density
limit,
the potential of mean force
$\psi_{12}(r)$ between two ions of respective charges $e_1$ and $e_2$ 
behaves like
$e_1e_2\exp(-\kappa r)/r$ as $r\to \infty$ rather  than like the Coulomb
potential itself $e_1e_2/r$, where $\kappa=(4\pi\beta \sum_i \rho_i
e_i^2)^{1/2}$ is the inverse  Debye shielding length. Here $\beta
=1/kT$
($k$ Boltzmann constant, $T$ temperature), $\rho_i$ is the density of
ionic species $i$ and the dielectric constant $D$ of the solvent has
been 
absorbed  in the definition of the charges. An important consequence of
the 
shielding effect is the non-analyticity of the specific excess osmotic
free energy $f(\rho)$ 
as a function of the mean ionic density $\rho$.
 Actually, in Debye-H\"uckel theory one
obtains $f(\rho)\sim \kappa^3 \sim \rho^{3/2}$ for $\rho \to 0$.

The results of Debye and H\"uckel are valid only at very low densities
and
discrepancies between their theory and experimental data on real
electrolytes
have motivated an enormous amount of theoretical works to improve the 
theoretical scheme. A first systematic pertubative expansion of $f(\rho)$
in
which the density $\rho$ is taken to be the ordering (small) parameter
was
proposed by Mayer. \cite{Mayer} Improvements on this seminal work were
made
later by Haga, \cite{Haga} Meeron, \cite{Meeron}Abe, \cite{Abe} and
Friedman.
\cite{Friedman0} All these works are based on  diagrammatic
techniques and
the more elaborate of them 
provide a reasonably accurate description of the thermodynamic
properties of
ionic solutions in the low fugacity regime. A monography by Friedman
\cite{Friedman} summarizes the above-mentioned works.

In a more recent work, \cite{Stell} Stell and Lebowitz have proposed a
perturbation scheme in which the ordering  parameter is
$\gamma=\beta e^2$ where $e$ is the electron charge.
Their theory is also based on a sophisticated diagrammatic
analysis
which gives an explicit high temperature expansion of $f(\rho)$ 
for symmetric and asymetric electrolytes.

Edwards \cite{Edwards} seems to have been the first to use 
the so-called sine-Gordon (SG)
transformation in the field of the statistical mechanics of
classical Coulomb systems as an alternative to the above-mentioned
diagrammatic techniques although Kac,\cite{Kac}
Siegert,\cite{Siegert} Hubbard, \cite{Hubbard} and Stratonovich
\cite{Stratonovich} also pioneered the method in other domains of
statistical physics or in field theory. In his work, Edwards considers a
model of
charged hard spheres which allows a clear splitting of   the pair
potential 
into a long range electrostatic part for which the sine-Gordon 
transformation
applies and a repulsive part for which low density virial expansion
techniques
can be employed. This leads to an intricate double expansion in 
$\gamma$ and in $\rho$. The sine-Gordon formalism for both classical
and
quantum Coulomb systems has been reviewed recently by Brydges and
Martin.\cite{Martin}

A decade later, Hubbard and Schofield\cite{Schofield} have shown that 
a general fluid Hamiltonian with long-range and short-range
interactions can be mapped onto a reference system with short-range interactions 
only. Then the cumulant expansion is used to map the original fluid Hamiltonian
onto a magneticlike Hamiltonian. Brilliantov and al.\cite{Brilliantov} have
explored this route for ionic fluids in order to study the Coulombic criticality.
The equation of state of a multicomponent system of {\em pointlike} ions embedded
 in a neutralizing background has also been studied by Ortner\cite{Ortner}
along these lines.
Recently Netz and Orland\cite{Netz} tried to improve on Edwards theory
by
performing a double SG transform, both on the Coulomb and the hard core
parts of the pair potential. As stressed by Brydges and Martin\cite{Martin}
a SG transform of the singular
hard core potential and more generally that of a repulsive short range
potential such that $\sim 1/r^{12}$,
is strictly speaking impossible, since these singular potentials do not have a 
Fourier  transform. 

The present work is along the lines of the papers of Brilliantov,Ortner, Netz and 
Orland.
We limit ourselves to the case of a symmetric 
fluid of charged hard spheres with only two
species of ions of equal  diameters $\sigma$ and carrying opposite charges
 (the so-called restrictive primitive model (RPM) of electrolytes\cite{Mayer}).
In a first step, we regularize the Coulomb potential by a smearing
of the
charges  over the surface of a sphere of diameter $a\leq \sigma$, and
therefore give a precise meaning to the SG transformation.\cite{Martin,Samuel}
Obviously other kinds of smearing are possible and would lead to the
same results; in another context an uniform volumic smearing of the charge has
been proposed.\cite{Ha} 
This allows us to derive
rigorously a result which seems to belong to 
Siegert\cite{Siegert} in the general case and which
states that the grand-partition function 
of charged hard spheres is equal to the
average over a Gaussian measure of the grand-partition function
$\Xi_{HS}$
of bare hard
spheres in the sine-Gordon field. This is our Eq.\ (\ref{eqsieg}).
In a second step, making a connection with liquid theory,\cite{Hansen2}
we perform a functional expansion of $\ln \Xi_{HS}$ with
respect to the sine-Gordon field $\phi$ 
which yields the exact expression\ (\ref{mainresu})
of the sine-Gordon action ${\cal
S}(\phi)$ of the model. 
This action involves the connected correlation functions of the hard
sphere fluids which are supposed to be known.
It can be checked that this action reduces to the usual sine-Gordon action
in the limit of vanishing hard-core diameters. 
This formalism is very
handy since
it allows to obtain either low fugacity or high temperature expansions
of $f(\rho)$ via cumulant expansions for off-lattice Coulombic systems.

The paper is organized as follows. In Sec. \ref{model}, we derive the generalized
sine-Gordon representation of the grand-partition function of the RPM model. The
low-density expansion of the grand potential (or pressure) of the model is obtained
in Sec. \ref{cumu} up to the order $\rho^{5/2}$.
 We check  that each term of this expansion is actually independent 
of the smearing diameter $a$. Comparisons  with 
the results of Haga and of Netz-Orland are carried out in Sec. \ref{discuss}.
All approaches yield identical results at order $\rho^{5/2}$. In addition 
we explain 
why the approximate derivation of Netz-Orland leads to the correct
result at this order but could fail at higher orders in $\rho$. Conclusions
are drawn in Sec. V.
%%%%%%%%%%%%%%%%%%%%%%%%%%%%%%%%%%%%%%%%%%%%%%%%%%%%%%%%%%%%%%%%%%%%%%%%%%%%%%%%
\section{Model and Formalism}
\label{model}
\subsection{The Boltzmann factor}
Throughout this paper we consider the three dimensional ($3D$) and
symmetric 
version of the RPM,  i.e.  a system made of 
$N_{+}$ hard spheres of diameter $\sigma$ and charge $e$ and $N_{-}$
spheres of
the same diameter but with an opposite charge $-e$.\cite{Friedman}

With obvious notations, the configurational energy of the model reads as
\begin{eqnarray}
\label{confi}
\beta V(\vec{r}\/ \/^{N_+},\vec{r}\/ \/^{N_-}) &=& \frac{\beta }{2}
{\sum_{i\neq j}} \left( \frac{e_i e_j}{r_{ij}}   
+v_{hs}(r_{ij}) \right) \; , 
 \end{eqnarray}
where $e_i=\pm e $ and $v_{hs}(r)$
denotes the hard core potential. For the moment the ions are supposed to
be
confined in some arbitrary volume $V \subset \R^3$.

We first note that only configurations $(\vec{r}\/ \/^{N_+},\vec{r}\/
\/^{N_-})$
of ions without overlaps of the spheres
do contribute to the canonical (or grand-canonical) partition functions.
 For these configurations,
the charge $\pm e$ 
of any ion of center $\vec{r_i}$ can be
smeared out uniformly on any spherical surface of diameter $0<a \leq \sigma$. 
The interaction energy of two balls of charge density 
$\tau(r)=\delta(r-a/2)/(\pi a^2)$, located respectively
at point $\vec{r}$ and $\vec{r}\; '$ will be noted $W_{\tau}(\vec{r} -
\vec{r}\; ')$ 
and we have obviously

\begin{eqnarray}
\label{Wtau}
W_{\tau}(\vec{r} - \vec{r}\;' )&=&  \int_{V} \int_{V} d^3\vec{x}\;
d^3\vec{y} \;
\tau(\vert \vec{r} - \vec{x} \vert) \frac{1}{\vert \vec{x}-\vec{y}
\vert}
 \tau(\vert
\vec{y} - \vec{r}\;' \vert) \; . 
\end{eqnarray}

We note that the self-energy  ${\cal E}_S \equiv W_{\tau}(0)/2=1/a$ of each spherical
distribution is a finite quantity for $a>0$.

It follows from the preceding remarks that the electrostatic part of the Boltzmann factor
can be written, for any configuration
 $ \; \; (\vec{r}\/ \/^{N_+},\vec{r}\/ \/^{N_-})$, as 

\begin{equation}
\label{Boltz}
\exp \left ( -\frac{\beta }{2}
{\sum_{i\neq j}} \frac{e_i e_j}{r_{ij}} \right )
=\exp \left(  \beta N \gamma {\cal E}_S \right)
\exp \left( -\frac{\gamma}{2} \langle n \vert W_{\tau} \vert n
\rangle \right)\; .
\end{equation}
where $\gamma \equiv \beta e^2$ and $N=N_+ + N_-$ is the total number of ions.
In Eq.\ (\ref{Boltz}) 
\begin{equation}
n(\vec{r}) \equiv \sum_{i=1}^{N_+} \delta(\vec{r} -
\vec{r}_{i+}) -  \sum_{i=1}^{N_-} \delta(\vec{r} - \vec{r}_{i-})
\end{equation}
is the microscopic charge distribution (divided by $e$) and 
\begin{equation}
\label{quadra}
\langle n \vert W_{\tau} \vert n \rangle \equiv
\int_{V} \int_{V} d^3\vec{r} \; d^3\vec{r}\; '
n(\vec{r})W_{\tau}(\vert \vec{r} - \vec{r}\; ' \vert)n(\vec{r}\; ') \; .
\end{equation}
Note that it follows from the positivity of the Fourier transform
$\tilde{W_{\tau}}(\vec{k})= 4\pi \tilde{\tau}(k)^2/k^2$ that the quadratic
form
$\langle n \vert W_{\tau} \vert n \rangle $ is definite positive. We can
take
advantage of this positivity to perform a SG transform\cite{Martin} and
reexpress the Boltzmann factor\ (\ref{Boltz}) as
an average over a Gaussian  scalar field $\phi$, i.e.

\begin{eqnarray}
\label{Wick}
\exp\left(-\frac{\gamma}{2} \langle n \vert W_{\tau} \vert n
\rangle
\right)&=&
\langle \exp \left(  i \int_V d^3  \vec{x} \; 
n(\vec{x})\overline{\phi}(\vec{x}) \right) \rangle_{W_{\tau}} 
= \langle \exp \left( i \sum_{i=1}^{N_+}\bar{\phi}(\vec{r}_{i+}) - i
\sum_{i=1}^{N_-}\bar{\phi}(\vec{r}_{i-}) \right) \rangle_{W_\tau} \; ,
\end{eqnarray}
where $\overline{\phi}(\vec{x}) \equiv \sqrt{\gamma} \phi(\vec{x})$
is a
real random field. The precise meaning of the average in Eq.\ (\ref{Wick}) 
is given in Appendix A.

\subsection{Grand partition function}
For reasons which should become clear below, the grand-canonical
ensemble is
considerably more handy than the canonical one. For simplicity we choose the same
chemical
potential $\mu \equiv \mu_+=\mu_-$ for the anions and the cations
\cite{Martin}. The
grand-canonical partition function is given by \cite{Hansen2}

 \begin{eqnarray}
\label{grandpa}
 \Xi^{RPM}(\nu_+,\nu_-) 
 \equiv  \sum_{N_+=0}^\infty \sum_{N_-=0}^\infty
 \frac{z^{N_+}}{(N_+)!} \frac{z^{N_-}}{(N_-)!}
 \int_{V} d^3\vec{r}^{N_+} d^3\vec{r}^{N_-}
 \exp \left (- \beta V(\vec{r}^{N_+},\vec{r}^{N_-})
 \right)\; .
\end{eqnarray}

 where we have introduced the usual notation $\nu
\equiv
\nu_{\pm}=\beta \mu$
and the activity $z\equiv z_{\pm}=\Lambda^{-3}\exp(\beta
\mu)$ of both species (the 
 thermal length $\Lambda$ is
assumed to be 
the same for the anions and the cations).

Gathering the intermediate results\ (\ref{Boltz}), (\ref{Wick}) we get :
\begin{eqnarray}
\Xi^{RPM}(\nu_+,\nu_-) &=&  \sum_{N_+=0}^\infty \sum_{N_-=0}^\infty
 \frac{\bar{z}^{N_+}}{(N_+)!} \frac{\bar{z}^{N_-}}{(N_-)!} 
 \int_{V} d^3\vec{r}^{N_+} d^3\vec{r}^{N_-}
 \exp \left(- \sum_{i<j}v_{hs}(r_{ij})\right)\\
&\times&
\langle \exp \left(
+ i \sum_{i=1}^{N_+}\bar{\phi}(\vec{r}_{i+}) -
i\sum_{i=1}^{N_-}\bar{\phi}(\vec{r}_{i-}) \right)
\rangle_{W_\tau} \; .
 \end{eqnarray}
where we have 
defined
renormalized chemical potentials 
$\bar{\nu}_{\pm}\equiv \nu_{\pm}+\beta e^2 {\cal E}_S  = \nu +\gamma/a$
and renormalized activities
$\bar{z}\equiv\bar{z}_{\pm}=\exp(\bar{\nu}_{\pm})/\Lambda^3$.

This last equation can be elegantly rewritten as 

\begin{equation}
\label{eqsieg}
\Xi^{RPM}(\nu_+,\nu_-) = \langle
\Xi^{HS}_{+,-}(\bar{\nu}_+,\bar{\nu}_-;i\bar{\phi},
-i\bar{\phi})\rangle_{W_\tau} \; ,
\end{equation} 
where $\Xi^{HS}_{+,-}(\bar{\nu}_+,\bar{\nu}_-; i\bar{\phi},
-i\bar{\phi})$ denotes the
grand-canonical partition function of a mixture of two species of equal
size
hard spheres
labelled $+$ and $-$  with chemical
potentials $\bar{\nu}_+$ and $\bar{\nu}_-$ respectively.  
The spheres with the label $+$ are in the external field
$i\bar{\phi}$ whereas those labelled $-$ are in the field
$-i\bar{\phi}$. Eq.\ (\ref{eqsieg}) is a special case of a more general
result
due to Siegert.\cite{Siegert} 

In order to get a more explicit expression of the action we perform now
a Taylor functional expansion of $\ln \Xi^{HS}_{+,-}$ with respect to the
activity $\bar{z}_{\pm}(\vec{r}) \equiv z \exp(\gamma/a)\exp( \pm i
\bar{\phi}(\vec{r}))$.

We have, from standard liquid theory \cite{Hansen2}

\begin{eqnarray}
\label{log}
& &\ln
\left(\frac{\Xi^{HS}_{+,-}(\bar{\nu}_+,\bar{\nu}_-,V,\beta;i\bar{\phi},
-i\bar{\phi})}{\Xi^{HS}_{+,-}(\nu,\nu,V,\beta)}
\right) \nonumber \\
&=&\sum_{n=1}^{\infty} \frac{1}{n!}\sum_{\alpha_{1}\cdots\alpha_{n}=\pm}
\int_{V} d^3 1 \cdots d^3 n
 \left. \frac{\delta^n \ln \Xi^{HS}_{+,-}}
{\delta \bar{z}_{\alpha_1}(1)\cdots \delta \bar{z}_{\alpha_n}(n)}
 \right\vert_{\bar{z}_{\alpha_{i}}(i) = z}
\prod_{i=1}^n \left(\bar{z}_{\alpha_i}(i)-z \right) \; ,
\end{eqnarray}

In the absence of an external field, we have, for sufficiently
large
systems
$
\label{mix}
\Xi^{HS}_{+,-}(\nu,\nu)=\Xi^{HS}(\nu_0)
\; ,
$ 
where $\nu_0 \equiv \nu+\ln 2$ (i.e.
$z_0=2z$)
and $\Xi^{HS}(\nu_0)$ is the
grand partition function of a fluid of identical hard spheres.
The integral kernels in Eq.\ (\ref{log}) are related to the
correlation functions $h_0^{(n)}(1, \cdots, n)$
of the hard sphere mixture \cite{Hansen2}

\begin{equation}
\label{core}
z^n \left. \frac{\delta^n \ln \Xi^{HS}_{+,-}}
{\delta \bar{z}_{\alpha_1}(1)\cdots \delta \bar{z}_{\alpha_n}(n)}
  \right\vert_{\bar{z}_{\alpha_{i}}(i) = z}
=\frac{\rho_0^n}{2^n} h_0^{(n)}(1, \cdots, n) \;.
\end{equation}

 In Eq.\ (\ref{core}), $\rho_0
=V^{-1} \partial \ln \Xi^{HS}/ \partial \nu_0$ 
is the number density of this fluid. 
Making use of of the identity
\begin{equation}
\sum_{\alpha_i=\pm}\frac{\bar{z}_{\alpha_i}(i)-z}{2z}
=\exp \left(\gamma/a \right)\cos\bar{\phi}(i) - 1 
\end{equation}
we can rewrite Eq.\ (\ref{log}) as a functional integral

\begin{eqnarray}
\label{mainresu}
\frac{\Xi^{RPM}(\nu_+,\nu_-)}{\Xi^{HS} (\nu_0)} &=&
\langle \exp(-U[\phi]) \rangle_{W_\tau} =
{\cal N}^{-1}_W \int {\cal D}\phi \; \exp( -{\cal S}[\phi]) \; ,
\end{eqnarray}
where ${\cal N}_W \equiv  \int {\cal D}\phi \; 
\exp( -\frac{1}{2} \langle \phi \vert W_{\tau}^{-1} \vert \phi \rangle) $
 is a normalization constant and 
\begin{eqnarray}
{\cal S}[\phi]&=& \frac{1}{2} \langle \phi \vert W_{\tau}^{-1}
\vert \phi \rangle + U[\phi] \; ,\\
U[\phi]&=&\sum_{n=1}^{\infty} U_n[\phi] \;, \nonumber \\
U_n[\phi]&=&  -\frac{\rho_0^n}{n!}
\int_{V} d^3 1 \cdots d^3 n \;
 \bar{h}_0 (1, \cdots, n)
\prod_{i=1,n}\left[\exp(\gamma/a)\cos\bar{\phi}(i) - 1 \right] \; .
\label{expa} 
\end{eqnarray}
The above expression of the sine-Gordon like action ${\cal S}[\phi]$ of
the RPM
is an exact result;
note that ${\cal S}[\phi]$ is an even function of the field. In the
limit $\sigma \to
0$ only the term $n=1$ of Eq.\ (\ref{expa}) survives and one checks that
one
recovers the usual sine-Gordon action of the Coulomb gas.\cite{Martin} 

A similar result along the Hubbard-Schofield scheme has been used by 
Brilliantov and al.\cite{Brilliantov} in their study of the criticality
of the RPM model. However their approach is developed in Fourier space
and without the explicit regularization obtained with the smearing of the charges.
As a final remark, we note that we might have been tempted to perform the
Taylor functional expansion, not around $z$ but around $\bar{z}_\pm = z
\exp(\gamma/a)$. In that case a low-fugacity expansion is valid only 
at high temperatures since $\bar{z} \sim z$ 
requires $\gamma \to 0$.
%%%%%%%%%%%%%%%%%%%%%%%%%%%%%%%%%%%%%%%%%%%%%%%%%%%%%%%%%%%%%%%%%%%%%%
\subsection{The generalised screened potential $X_\tau(r)$}

The program is now to perform a systematic cumulant expansion of the
expression\ (\ref{mainresu}) and to compute the cumulants by an
extensive use of Wick's theorem.\cite{Binney,Ma}
However, as it stands, this expansion will
involve cumulants which diverge in the thermodynamic limit due to the
long range of $W_{\tau}(r)$. The same problem arises in low fugacity or
high
temperature diagrammatic expansions
of the RPM where one is led to resum classes of diagrams in order
to get finite results.\cite{Mayer,Stell,Hansen2}
 Formally it amounts to introduce a screened
(Yukawa) potential. In the field theoretical formalism discussed here,
 a screened (or
Hartree \cite{Parisi}) field can also be introduced as follows.
We denote $U_0[\phi]$ the high temperature approximation of $U[\phi]$,
i.e.
\begin{equation}
\label{U0}
U_0[\phi] = \frac{\rho_0}{2}  \int d^3\vec{r} \bar{\phi}(\vec{r})^2 \; .
\end{equation}
 Writing now the triviality 
$U[\phi]= \left(U[\phi]-U_0[\phi]\right) +U_0[\phi]$, we
get
\begin{eqnarray}
\label{EU}
 \langle \exp \left(-U[\phi]\right) \rangle_{W_\tau} &=& 
\frac{{\cal N}_X}{{\cal N}_W}
\langle \exp \left(- \left(U[\phi] -  U_0[\phi] \right) \right)\; ,
\end{eqnarray}
where ${\cal N}_A \equiv \int {\cal D}\phi \; 
\exp( -\frac{1}{2} \langle \phi \vert A_{\tau}^{-1} \vert \phi \rangle) $
with $A=X_\tau,W_\tau$ and where $X_\tau(r)$ is a real operator defined by the
 relation $X_\tau(r)^{-1}\equiv W_\tau(r)^{-1} + \rho_0 \gamma \;{\rm I}$ ({\rm I}
denotes the identity).

Using the precise definition of the functional integration given in Appendix A,
we obtain
\begin{eqnarray}
\frac{{\cal N}_X}{{\cal N}_W}
\label{x}
&=&  
\exp \left(- \frac{V}{2} \int \frac{d^3\vec{q}}{(2\pi)^3} \ln \left( 1
+ \gamma \rho_0 \tilde{W}_\tau(\vec{q}) \right) \right) \; ,
\end{eqnarray}

Note that, in Eq.\ (\ref{x}), we have replaced a series by an integral,
which is
valid for large systems. Moreover the integral converges for $a\neq 0$,
which is a happy consequence of the regularization of the Coulomb
potential via the smearing of the charge.

The Fourier transform $\tilde{X}_{\tau}(q)$ reads :
\begin{eqnarray}
\label{Xq}
\tilde{X}_{\tau}(q) &=&\frac{\tilde{W}_\tau(q) }
{1 +  \gamma \rho_0 \tilde{W}_{\tau}(q)}=
 \frac{\sin^2 (qa/2)}{(qa/2)^2}
\frac{4\pi}{q^2 +  \kappa_0^2 \frac{\sin^2(qa/2)}{(qa/2)^2}} \; . 
\end{eqnarray}
with $\kappa_0^2 = 4\pi \gamma \rho_0$.

It is easily checked that for a fixed $\kappa_0$ and 
in the limit $a\rightarrow 0$, $X_\tau(r)$ reduces to
the familiar screened Yukawa potential, i.e. 
$X_\tau(r) \sim \exp (-\kappa_0 r)/r$ ($\forall r$). 
Conversely, for a fixed $a$ and in the
limit $\kappa_0 \rightarrow 0$ we have obviously $X_{\tau}(r)= W_{\tau}(r)$.
For arbitrary $(a,\kappa_0)$, the large $r$ behavior of $X_{\tau}(r)$ is
determined
by the small $q$ behavior of $\tilde{X}_{\tau}(q)\sim 4 \pi/(q^2 
+\kappa_0^2)$
 which implies $X_{\tau}(r) \sim \exp(-\kappa_0 r)/r$ at large $r$. The
function
$X_{\tau}(r)$ is thus a short range function of $r$ which, however, for
large
$\kappa_0$, can have a non monotoneous behavior. Expansions of $X_\tau(r)$
 at low $\kappa_0$ will be given in Sec.\ \ref{cumu}.

 We shall see
in the next Section that the use of the screened
potential $X_{\tau}(r)$ instead of the long range (Coulombic) potential
$W_{\tau}(r)$ ensures the convergence of the cumulants.
%%%%%%%%%%%%%%%%%%%%%%%%%%%%%%%%%%%%%%%%%%%%%%%%%%%%%%%%%%%%%%%%%%%%%%%%%%%%%%%%
%%%%%%%%%%%%%%%%%%%%%%%%%%%%%%%%%%%%%%%%%%%%%%%%%%%%%%%%%%%%%%%%%%%%%%%%%%%%%%%%
\section{Low-fugacity expansion}
\label{cumu}

In this section, we use the general result of Eq.\ (\ref{mainresu}) to perform
a systematic low-fugacity expansion of the pressure of the RPM model,
 at the order $5/2$ in the density $\rho$.

Although this theory must not depend explicitely of the smearing parameter $a$,
we shall explicitely verify below that it is effectively the case for each
order of the expansion.

\subsection{The grand potential}

Using the results of Eqs.\ (\ref{mainresu}), (\ref{EU}), (\ref{x}),  and the 
cumulant theorem
\cite{Ma} we obtain the specific grand potential $ \omega_{RPM} \equiv 
-\ln \Xi_{RPM} /V$ 
\begin{eqnarray}
\label{omegagene}
\omega_{RPM}(\nu) &=& \omega_{HS}(\nu_0) 
 + \frac{1}{2}
\int \frac{d^3\vec{q}}{(2\pi)^3} \ln \left( 1 
+ \gamma \rho_0 \tilde{W}_\tau(q) \right) 
- \frac{1}{V}
\sum_{n=1}^{\infty} \frac{(-1)^n}{n!}
\langle {\cal H}^n[\phi]\rangle_{X_\tau,c}\; .
\end{eqnarray}
where ${\cal H}[\phi] \equiv U[\phi] - U_0[\phi]$ and $\langle \cdots \rangle_{X_\tau,c}$
denotes a cumulant average.

Since we are interested in a low-density expansion, we keep only the first two 
terms $U_1$ and $U_2$ of the series (\ref{expa}), which is equivalent to take
into account all contributions with one and two point correlations functions.
Then, ${\cal H}[\phi] = U_1[\phi]+U_2[\phi]-U_0[\phi]$, and the grand potentiel
can be recast in the following form :
\begin{eqnarray}
\omega_{RPM}(\nu) &=& \omega_{HS}(\nu_0) +\omega_1
+\omega_2 + {\cal O}(\rho_0^3)\; ,
\end{eqnarray}
with the following definitions :
\begin{eqnarray}
\label{omega1}
\omega_1 &\equiv&  \frac{1}{2}
\int \frac{d^3\vec{q}}{(2\pi)^3} \ln \left( 1 
+ \gamma \rho_0 \tilde{W}_\tau(q) \right) \; ,\\
\omega_2 &\equiv&  \frac{\langle {\cal H}\rangle_{X_\tau}}{V} 
- \frac{\langle {\cal H}^2\rangle_{X_\tau}-
\langle {\cal H}\rangle^2_{X_\tau}}{2V} \; .
\end{eqnarray}

The well-known $HS$ contribution is given by :
\begin{eqnarray}
\omega_{HS}(\nu_0)&=& -\rho_0 -\frac{2\pi}{3}\rho_0^2 \sigma^3 +
 {\cal O}(\rho_0^3)\; .
\end{eqnarray}

$\omega_1$ is a generalization, when smearing is taking
into account, of the familiar DH expression of the free energy.
 Notice that, unlike the point-like DH approach,
the integral\ (\ref{omega1}) which defines $\omega_1$
 is  convergent  at large $k$
 (within the conventional theory, an infinite self-energy 
must be substracted to recover finite results). Since it includes $\tilde{W}(q)$,
$\omega_1$ is a function of $a$ (and $\kappa_0$); its expansion
in powers of $\kappa_0$ is given in Appendix B and reads :
\begin{eqnarray}
\label{logarithme}
\frac{1}{2}\int \frac{d^3\vec{q}}{(2\pi)^3} \ln \left( 1 
+ \gamma \rho_0 \tilde{W}_\tau(q) \right) -
\frac{\rho_0 \gamma}{a} = -\frac{2\sqrt{\pi}}{3}\rho_0^{3/2}\gamma^{3/2}
+ \frac{7\pi a}{15} \rho_0^2 \gamma^2 - \frac{\pi \sqrt \pi a^2}{3}
\rho_0^{5/2}\gamma^{5/2}+ {\cal O}(\rho_0^3)\; .
\end{eqnarray}
The first term recast in the form $-\kappa_0^3/12\pi$ is reminiscent of 
the familiar DH contribution to the free energy.

$\omega_2$ contains averages over the gaussian field 
$\bar{\phi}$ which can be obtained from application of 
Wick's theorem; the detailed calculation is reported in Appendix C and one finds
\begin{eqnarray}
\label{om2}
\omega_2 &=& -\frac{\gamma \rho_0}{a} -\frac{\rho_0 \Delta_0}{2}
+\rho_0 \left[1 -\exp(-\Delta_0/2 )\right]
-\frac{\rho_0^2}{4}\left[1 -\exp(-\Delta_0/2 )\right]^2
\int d\vec{r} \psi_\tau^2(r)\\ \nonumber
&-&\frac{\rho_0^2}{2}
\left[1-\exp(-\Delta_0/2)\right]^2\tilde{h}_0(0)
+ \frac{\rho_0^2}{4}\exp(-\Delta_0)\int_{r<\sigma} d\vec{r} \psi_\tau^2(r)
-\frac{\rho_0^2}{2}\exp(-\Delta_0)
\sum_{n=2}^\infty 
\int_{r>\sigma} d\vec{r} \frac{\psi_\tau^{2n}(r)}{(2n)!}\; .
\end{eqnarray}
where $\tilde{h}_0(0)$ denotes the 3D Fourier transform of $h_0^{(2)}(r)$.
We have introduced in this last expression the  dimensionless potential $\psi_\tau(r)
\equiv \gamma X_\tau(r)$ and the quantity $\Delta_0 \equiv \psi_\tau(0) -\frac{2\gamma}{a}$.
Expansions of $\psi_\tau(r)$ and $\Delta_0$ at low $\kappa_0$ are given in Appendix B
\begin{eqnarray}
\label{D0}
\Delta_0 &\equiv& \psi_\tau(0)-\frac{2\gamma}{a}= \gamma \left[
-\kappa_0 +\frac{7}{15}\kappa_0^2 a -\frac{5}{24}\kappa_0^3 a^2 +
{\cal O}(\kappa_0^4) \right]\; ,
\end{eqnarray}
while
\begin{eqnarray}
\label{psi<}
\psi_\tau(r)&=&\frac{2\gamma}{a}-\frac{\gamma}{a^2}r-\gamma \kappa_0 
+ {\cal O}(\kappa_0^2) \quad \mbox{for} \quad r < a\; ,\\
\label{psi>}
\psi_\tau(r)&=& \gamma q_0^2 \frac{\exp(-\kappa_0 r)}{r}
+ {\cal O}(\kappa_0^2) \quad \mbox{for} \quad r > a\; .
\end{eqnarray}
with $q_0= 2 \sinh(\kappa_0 a /2)/(\kappa_0 a)$.

Using (\ref{D0}),(\ref{psi<}) and (\ref{psi>}) each contribution to  (\ref{om2})
 can be easily expanded in powers of $\rho_0$.
 The first two terms give
\begin{eqnarray}
\label{o21}
-\frac{\rho_0 \Delta_0}{2}
+\rho_0 \left[1 -\exp(-\Delta_0/2 )\right]&=&
-\frac{\pi}{2}\rho_0^2 \gamma^3
-\frac{\pi \sqrt \pi \gamma^{9/2} \rho_0^{5/2}}{6}
+ \frac{14 \pi \sqrt \pi \gamma^{7/2} \rho_0^{5/2}a}{15}
 +{\cal O}(\rho_0^3)\; .
\end{eqnarray}

The prefactor of the third term is $\rho_0^2\left[1-\exp(-\Delta_0/2)\right]^2 =
 \pi \gamma^3 \rho_0^3 +{\cal O}(\rho_0^{7/2})$ and the integral must be
 splitted 
into two components :
\begin{eqnarray}
\label{o22}
\int_{r<\sigma} d\vec{r} \psi_\tau^2(r)&=&
\left(\frac{32 \pi \gamma^2 a}{15}- \frac{20 \pi \sqrt \pi \gamma^{5/2} \rho_0^{1/2} a^2}{3}
 \right) \nonumber\\
&+&4\pi \gamma^2 (\sigma -a) + 8 \pi \sqrt \pi \gamma^{5/2} \rho_0^{1/2}(a^2 -\sigma^2)
 +{\cal O}(\rho_0)\; , \\
\int_{r>\sigma} d\vec{r} \psi_\tau^2(r)&=&
\frac{\sqrt \pi \gamma^{3/2}}{ \rho_0^{1/2}}+{\cal O}(1)\; . 
\end{eqnarray}
Notice that the last integral is singular in $\rho_0$ when $\rho_0 \to 0$. Consequently
the term under investigation will contribute at the order $\rho_0^{5/2}$.

The expansion of the prefactor $\rho_0^2 \exp(-\Delta_0)$ is straightforward
$\rho_0^2 \exp(-\Delta_0)=
\rho_0^2 + 2\sqrt \pi \gamma^{3/2} \rho_0^{5/2}+ {\cal O}(\rho_0^3)$, while
$\int_{r>\sigma} d\vec{r} \psi_\tau^{2n}(r)$ for $n \geq 2$ is related to the 
exponential integral function ${\rm E}_n(z)= \int_1^\infty dt \exp(- zt) t^{-n}$.
Indeed one finds
\begin{eqnarray}
\label{o23}
\sum_{n=2}^{\infty}\int_{r>\sigma} d\vec{r} \frac{\psi_\tau^{2n}(r)}{(2n)!}&=&
4\pi \sigma^3 \sum_{n=2}^{\infty} \frac{(\gamma/\sigma)^{2n}q_0^{4n}}{(2n)!}
{\rm E}_{2n-2}(2n\kappa_0 \sigma)\; \nonumber \\
&=&
4\pi \sigma^3\sum_{n=2}^{\infty}\frac{(\gamma/\sigma)^{2n}}{(2n)!(2n-3)}
+\gamma^4\frac{2\pi \kappa_0 }{3}
\left(\gamma_E + \ln(4\kappa_0 \sigma)-1\right)\nonumber \\
&-& 4\pi \sigma^3 \kappa_0 \gamma
\sum_{n=2}^{\infty}\frac{(\gamma/\sigma)^{2n+1}}{(2n+1)!(2n-2)}
+{\cal O}(\rho_0)
\end{eqnarray}
where the last equality is obtained with the help of the series representation of 
the exponential integral function\cite{Abra}.

Gathering all these results, it can be checked that
all contributions involving the smearing diameter $a$  exactly cancel each other
and that we indeed obtain a result independent on $a$; more precisely we get 
\begin{eqnarray}
\label{omnu}
\omega_{RPM}(\nu)  &=& -\rho_0
 -\frac{2\sqrt{\pi}}{3}\rho_0^{3/2}\gamma^{3/2}
-  \rho_0^2 \left(\frac{2 \pi \sigma^3}{3}
-  \pi \gamma^2 \sigma
+  \frac{\pi}{2} \gamma^3
+ 2\pi \sigma^3 \; {\rm S}(\gamma/\sigma)\right) \nonumber \\
&-& \frac{\pi \sqrt \pi \gamma^{9/2} \rho_0^{5/2}}{12}
\left[8\gamma_E -3  + 8\ln(8 \sqrt \pi \gamma^{1/2} \sigma)\right]
+2\pi \sqrt \pi \gamma^{7/2} \rho_0^{5/2}\sigma
-2\pi \sqrt \pi \gamma^{5/2} \rho_0^{5/2}\sigma^2\nonumber \\
&-&4\pi \sqrt \pi \gamma^{3/2} \rho_0^{5/2}\sigma^3 \;
\left[ {\rm S}(\gamma/\sigma)- {\rm T}(\gamma/\sigma)\right]
- \frac{\pi \sqrt \pi \gamma^{9/2}}{3}\rho_0^{5/2}\ln(\rho_0)
+  {\cal O}(\rho^3)\; .
\end{eqnarray}
where we have introduced the two following series 
\begin{eqnarray}
\label{series}
{\rm S}(\gamma/\sigma) &\equiv &
\sum_{n=2}^{\infty}  \frac{(\gamma/\sigma)^{2n}}{(2n)!(2n-3)}\; ,\nonumber \\
{\rm T}(\gamma/\sigma)&\equiv& 
\sum_{n=2}^{\infty}\frac{(\gamma/\sigma)^{2n+1}}{(2n+1)!(2n-2)}\; .
\end{eqnarray}

%%%%%%%%%%%%%%%%%%%%%%%%%%%%%%%%%%%%%%%%%%%%%%%%%%%%%%%%%%%%%%%%%%%%%%%%%%%%%%%%
\subsection{Pressure of the RPM model}

In this section we derive the pressure for low
density systems at arbitrary temperatures. In order to obtain these quantities,
we must first performed a transformation from  the density $\rho_0$
(corresponding to the chemical potential $\nu_0$) to the activity $z$ of the
model. For that purpose we use the relation :
\begin{eqnarray}
\label{rtoz}
\rho_0 &=&- z_0\frac{\partial \omega_{HS}(\nu_0)}{\partial z_0} 
=z_0 -  \frac{4 \pi \sigma^3}{3}z_0^2 + {\cal O}(z_0^3)
= 2z- \frac{16 \pi \sigma^3}{3}z^2 + {\cal O}(z^3)
\; .
\end{eqnarray}

Inserting (\ref{rtoz}) in (\ref{omnu}) we get the expression of $\omega_{RPM}$
as a function of $z$ and $\gamma$
\begin{eqnarray}
\label{omz}
\omega_{RPM}(z,\gamma) &=& -2z
-\frac{4 \sqrt{2\pi} \gamma^{3/2}}{3}z^{3/2}
+4 \pi \sigma^3\left(\frac{2}{3}+  \left(\frac{\gamma}{\sigma}\right)^2
-\frac{1}{2}\left(\frac{\gamma}{\sigma}\right)^3
-2 \; {\rm S}(\gamma/\sigma)\right)z^2 
\nonumber \\
&+&z^{5/2}  4 \sqrt 2 \pi^{3/2} \left(
-\frac{ \gamma^{9/2}}{12}(8\gamma_E +8 \ln(8\sqrt \pi \gamma^{1/2}\sigma)-3)
+2  \gamma^{7/2} \sigma 
-2  \gamma^{5/2} \sigma^2 \right) \nonumber \\
&+& z^{5/2} \frac{16 \sqrt 2 \pi^{3/2} \gamma^{3/2}\sigma^3}{3}
\left(1-3\left[ {\rm S}(\gamma/\sigma)- {\rm T}(\gamma/\sigma)\right]
\right)
-z^{5/2}\ln(2z) \frac{4 \sqrt 2 \pi^{3/2} \gamma^{9/2}}{3}
+  {\cal O}(z^3)\; .
\end{eqnarray}

The density of the system is obtained by the relation 
$\rho = - z \partial \omega_{RPM}(z)/ \partial z $ which is easily inverted 
and reads
\begin{eqnarray}
z&=& \frac{\rho_0}{2}
-\frac{ \sqrt{\pi} \gamma^{3/2}}{2}\rho_0^{3/2}
+ \pi \sigma^3\left(\frac{2}{3}+  \left(\frac{\gamma}{\sigma}\right)^2
+\frac{1}{4}\left(\frac{\gamma}{\sigma}\right)^3
-2 \; {\rm S}(\gamma/\sigma)\right)\rho_0^2 
\nonumber \\
&+&\rho_0^{5/2}  \pi^{3/2} \left(
-\frac{ \gamma^{9/2}}{12}(10 \gamma_E +10 \ln(8\sqrt \pi \gamma^{1/2}\sigma)-7)
- \gamma^{7/2} \sigma 
- \frac{5 \gamma^{5/2} \sigma^2}{2} \right) \nonumber \\
&+& \rho_0^{5/2} \pi^{3/2} \gamma^{3/2} \sigma^3
\left(-\frac{2}{3}+2 {\rm S}(\gamma/\sigma)+5 {\rm T}(\gamma/\sigma)
\right)
-\rho_0^{5/2}\ln(\rho_0) \frac{5 \pi^{3/2} \gamma^{9/2}}{12}
+  {\cal O}(\rho_0^3)\; .
\end{eqnarray}

Using this last equation in  (\ref{omz}) we obtain the pressure
 $\beta P_{RPM} \equiv - \omega_{RPM}(\rho,\gamma) $ of the RPM model 

\begin{eqnarray}
\beta P_{RPM} & =&
 \rho -\frac{ \sqrt{\pi} \gamma^{3/2}}{3}\rho^{3/2}
+\sigma^3\left(\frac{2\pi}{3}+ \pi \left(\frac{\gamma}{\sigma}\right)^2
-2\pi \; {\rm S}(\gamma/\sigma)\right)\rho^2 \nonumber \\
&-&\left(
\pi \sqrt \pi \gamma^{9/2}
 \left(-\frac{2}{3}+\gamma_E +\ln(8\sqrt \pi \gamma^{1/2}\rho^{1/2}\sigma)\right)
+ 3\pi \sqrt \pi \gamma^{5/2} \sigma^2
- 6\pi \sqrt \pi \gamma^{3/2} \sigma^3 {\rm T}(\gamma/\sigma)
\right)\rho^{5/2} \nonumber \\
&+&{\cal O}(\rho^{3})
\end{eqnarray}
which can be recasted in the following form   :
\begin{eqnarray}
\label{pression}
\beta P_{RPM} &=& 
 \rho -\frac{\kappa^3}{24 \pi}
+\frac{2\pi\sigma^3}{3}\rho^2+ \frac{\kappa^4 \sigma}{16 \pi}
-2\pi \rho^2 \sigma^3 \; {\rm S}(\gamma/\sigma)
-\frac{\kappa \kappa_2^4}{512 \pi^3}
\left(-\frac{2}{3}+\gamma_E +\ln(4 \kappa \sigma)\right)
- \frac{3\kappa^5 \sigma^2}{32 \pi} \nonumber \\
&+& \frac{3}{4}(4\pi \gamma)^{3/2} \sigma^3 \rho^{5/2} {\rm T}(\gamma/\sigma)
+{\cal O}(\rho^{3})
\end{eqnarray}
with $\kappa_2^2 \equiv 4\pi \gamma \kappa^2$ and
$\kappa^2 = 4\pi \gamma \rho$, the square of the usual inverse DH length.
${\rm S}(\gamma/\sigma)$ and ${\rm T}(\gamma/\sigma)$ are defined by Eq.\ 
(\ref{series}).

%%%%%%%%%%%%%%%%%%%%%%%%%%%%%%%%%%%%%%%%%%%%%%%%%%%%%%%%%%%%%%%%%%%%%%%%%%%%%%%%
\section{Discussion}
\label{discuss}

In this section we briefly compare the results obtained above with the classical
 diagrammatic
results of Haga\cite{Haga} and with the field theoretical approach of Netz-Orland
\cite{Netz}. It will be shown below, as expected, that these three different 
routes yield the same result.

 Haga's expression for the equation of state (Eq.\ (29) of ref\cite{Haga})
 is easily compared with Eq.\
(\ref{pression}); both results coincide except for the terms proportionnal
to $\kappa^5$ which differ by a factor of $1/2$. In a recent paper by
Bekiranov and Fisher\cite{Bekiranov}, a slip in Haga's Eq.\ (25.4)
was noted by these authors who pointed out that the last term of Haga's equation
 should read $\kappa^5a^2/16\pi$ instead of $\kappa^5a^2/32\pi$. When this 
 correction
is taken into account  Haga's results  and ours
are identical.

In their paper\cite{Netz}, Netz and Orland do not compare explicitely their results 
with those of Haga. This can be  done by confronting our expression
of the grand potential in terms of the fugacity (cf Eq.\ (\ref{omz}))  
with Eq.\ (22) of ref\cite{Netz}. Their results are given in terms of the hyperbolic 
sine-integral function ${\rm Shi}(\gamma/\sigma)$ and of the incomplete Gamma
function $\Gamma(0,\gamma/\sigma)$.
As shown in Appendix\ D these functions can be reexpressed 
in terms of the series
${\rm S}(\gamma/\sigma)$ and ${\rm T}(\gamma/\sigma)$ defined by Eq.\ 
(\ref{series}) and it appears
that the two results coincide. Let us now explain, why the approximate
theory of Netz and Orland gives the correct result at this order. Recall that
these authors also perform  a
Hubbard-Stratonovich transformation on the singular
hard-core potential. The associate random field is denoted $\psi(r)$ and
the following average is needed (Eq.\ (12) of ref\cite{Netz})
\begin{equation}
\label{hh}
\langle h(1)h(2) \rangle = \exp(-w(12))
\end{equation}
where $h(r)\equiv \exp(-i\psi(r)+w(0)/2)$ and $w(r)$ denotes the hard-core potential.
Two consequences result from Eq.\ (\ref{hh}). On the
one hand, as noted by the authors,
$ \exp(-w(12))$ (contrary to $w$) is finite, which regularizes the theory. On the other
hand, since $\exp(-w(r))=1-\theta(r-a)$ ($\theta$ is the Heaviside function), this procedure
amounts to incorpore hard-core effects only at the level of the second virial
coefficient. It appears (see Eqs.\ (\ref{g1},\ref{g2}) of Appendix C), that this approximation is
sufficient for an expansion up to $\rho^{5/2}$ but will miss some contributions 
at the next order.

%%%%%%%%%%%%%%%%%%%%%%%%%%%%%%%%%%%%%%%%%%%%%%%%%%%%%%%%%%%%%%%%%%%%%%%%%%%%%%%%
\section{Conclusion}

In conclusion we have proposed in this work a formally exact field theory for
hard-core Coulomb systems. This approach generalizes the usual sine-Gordon
theory valid for pointlike charges to realistic Coulomb fluids with additional 
short range interactions.
Within this formalism we derive the equation of state
of the RPM model up to $\rho^{5/2}$. Our results confirm the classical diagrammatic 
expansions of the Mayer-Haga diagrammatic theory. Going to next order
 is perhaps not out of reach although there is a delicate analysis of the relevant
 contributions to the cumulants to perform. Note that an equation of state for
pointlike ions as been recently obtained by Ortner\cite{Ortner} using the 
Hubbard-Schofield approach, up to the $\rho^3$ contribution. 
In this latter case, however, 
the reference system is the ideal gas system which, in turn, greatly simplify the
 calculations.
In the companion paper, we derive an equation of state of the RPM in the high
temperature regime, at any density, by using the formalism developed
in the present paper. This two complementary 
limits show the ability of our formulation to tackle in a coherent way the equation
of state of Coulomb systems.

Another problem, which remains a challenge to theory, is the understanding of
ionic criticality\cite{Fisher}.  We believe
that our formalism might be used to give some insights upon these interesting 
questions. Work in that direction is currently in progress.
%%%%%%%%%%%%%%%%%%%%%%%%%%%%%%%%%%%%%%%%%%%%%%%%%%%%%%%%%%%%%%%%%%%%%%%%%%%%%%%%
\acknowledgments
The authors wish to acknowledge helpul comments of B. Jancovici, D. Levesque and
J. J. Weis. 
JLR thanks  the complex fluid group in CRMD (Orl\'eans) for stimulating
interactions.

%%%%%%%%%%%%%%%%%%%%%%%%%%%%%%%%%%%%%%%%%%%%%%%%%%%%%%%%%%%%%%%%%%%%%%%%%%%%%%%%
\appendix
\section{Functional integration}

For any operator of the real field $\phi(\vec{r})$ we define the average
$\langle A[\phi] \rangle_{W_\tau}$ by the following relation 
\begin{equation}
\langle A[\phi] \rangle_{W_{\tau}} =
\int {\cal D} \phi \; P_{W_{\tau}} [\phi] A[\phi] \; ,
\end{equation}
where $\int {\cal D}\phi $ denotes a functional integration and
the Gaussian weight $P_{W_{\tau}} [\phi]$ is defined as

\begin{equation}
P_{W_{\tau}} [\phi]=\frac{\exp\left(-\frac{1}{2}\langle \phi \vert
W_{\tau}^{-1}\vert \phi \rangle \right)}
{\int {\cal D}\phi \; \exp 
\left(-\frac{1}{2}\langle \phi \vert
W_{\tau}^{-1}\vert \phi \rangle \right)} \; .
\end{equation}
 In order to give an unambiguous definition of the measure ${\cal
D}\phi$ and
thus a precise meaning to the SG transform we
henceforth consider a cubic volume $V=L^3$ with periodic boundary
conditions
(PBC).
 The $1/r$ Coulomb potential which enters the configurational energy\
(\ref{confi}) of the RPM
must therefore be replaced by the Ewald potential
\cite{Teller,Hansen,Perram,Caillol}
\begin{equation}
\label{Ewald}
E(\vec{r})= \frac{4\pi}{L^3}\sum_{\vec{q}\neq \vec{0}}
\frac{\exp(i\vec{q}\cdot\vec{r})}{\vec{q}^{\;2}} \; ,
\end{equation}
where $\vec{q}=2\pi\vec{n} /L$ ($\vec{n} \equiv (n_x,n_y,n_z)  \in
\Z^3$) is a
vector of the reciprocal lattice.
Recall that $E(\vec{r})$ is
the periodic electrostatic potential of a point charge embedded in a
uniform 
neutralizing background which kills the term $\vec{q}=\vec{0}$ in the
series\
(\ref{Ewald}).\cite{Caillol,Felder} This causes a (hopefully) slight
difficulty 
since
configurations
with $N_+ \neq N_-$ are associated with the presence of a background
which
ensures the electric neutrality of the system. The periodical
system considered here is therefore slightly different from the usual
RPM;
however, this should make no difference in the thermodynamic limit.

Assuming PBC we thus have\cite{Binney}
\begin{equation}
\label{dphi}
\int {\cal D} \phi \equiv {\prod}'_{\vec{q}} 
\int_{-\infty}^{+\infty} d\tilde{\phi}_{\vec{q}}^R  
\int_{-\infty}^{+\infty}d\tilde{\phi}_{\vec{q}}^I \; ,
\end{equation}
where $\tilde{\phi}_{\vec{q}}^R$ and $\tilde{\phi}_{\vec{q}}^I$ denote
respectively
the real and imaginary parts of the Fourier component 

\begin{equation}
\tilde{\phi}_{\vec{q}}=\int_V 
d^3\vec{r}\;\phi(\vec{r})\exp(-i\vec{q}\cdot\vec{r}) 
\end{equation}
of the real field $\phi$.
 The infinite product in Eq.\ (\ref{dphi}) runs over
the vectors $\vec{q}\neq \vec{0}$ of the reciprocal 
lattice. In fact, since, due to the reality of the field $\phi$,
$\tilde{\phi}_{\vec{q}}=\tilde{\phi}_{-\vec{q}}^*$, only half of the
vectors
has to be considered, for instance those with $n_x \geq 0$.\cite{Binney}
That is what is meant by the subscript ' in Eq.\ (\ref{dphi}). 

%%%%%%%%%%%%%%%%%%%%%%%%%%%%%%%%%%%%%%%%%%%%%%%%%%%%%%%%%%%%%%%%%%%%%%%%%%%%%%%%
\section{Low $\kappa_0$ expansions}

In this Appendix, we give the calculation of the expansions
of $\Delta_0$, $\psi_\tau(r)$ and $\omega_1$ with respect to $\kappa_0$.

\begin{enumerate}
\item
Let us begin with $\Delta_0$; we recall that $\Delta_0 \equiv \psi_\tau(0) 
-2 \gamma/a $ where $\psi_\tau(0)=\gamma X_\tau(0)$. Using Eq.\ (\ref{Xq})
we get 

\begin{eqnarray}
\label{X0xi}
X_\tau(0) &=& \int \frac{d^3 \vec{q}}{(2\pi)^3} 
\tilde{X}_{\tau}(q) =
\frac{2}{\pi a } \int_{-\infty}^{+\infty}  
\frac{\sin^2 x}{x^2 +\xi_0^2 \frac{\sin^2 x}{x^2}}dx 
= \frac{2}{\pi a } \int_{-\infty}^{+\infty} 
\frac{\sin^2x}{x^2 +\xi_0^2}\; \frac{1}{1+ \nu(x)}\; dx \; ,
\end{eqnarray}
where the dimensionless parameter $\xi_0 = \kappa_0 a/2$ and the function
\begin{equation}
\nu(x)= \frac{\xi_0^2}{x^2 +\xi_0^2}\left[ \frac{\sin^2x}{x^2}-1 \right] \; 
\end{equation}
satisfies $\vert \nu(x) \vert < 1 \; \forall x$. Therefore the fraction
$1/(1+\nu(x))$ in Eq.\ (\ref{X0xi}) 
can be replaced by its series representation which yields
\begin{eqnarray}
\label{calF}
X_\tau(0) &=&  
 \frac{2}{\pi a}\sum_{n=0}^\infty (-)^n  {\cal I}_n(\xi_0) \; ,\nonumber
 \\
 {\cal I}_n(\xi_0) &=& 
 \int_{-\infty}^{+\infty} \frac{\sin^2 x}{x^2 +\xi_0^2} 
 \nu^n(x) dx \; , 
\end{eqnarray}
where it should be noted that the integrals ${\cal I}_n(\xi_0) $ in  Eq.\
(\ref{calF}) are entire functions of  $\xi_0$.
Consequently we get at order $\xi^3_0$
\begin{eqnarray}
\label{F0}
X_\tau(0) &=& \frac{2}{\pi a}
\left[{\cal I}_0(\xi_0) - {\cal I}_1(\xi_0) \right]
 +{\cal O}(\xi_0^4)                        \nonumber \; ,
\end{eqnarray}
 where ${\cal I}_0(\xi_0)$ and ${\cal I}_1(\xi_0)$ can be computed by means of
 the residue theorem
\begin{eqnarray}
{\cal I}_0(\xi_0)&=&\frac{\pi}{2\xi_0}\left(1-\exp(-2\xi_0)\right)\; ,\nonumber \\
{\cal I}_1(\xi_0)&=& \frac{\pi}{16 \xi_0^3}\left(-9 +
8 \xi_0 + (12 + 8\xi_0) \exp(-2\xi_0) - (3  + 4\xi_0 ) \exp(-4\xi_0)  \right)
\nonumber \\
&-&\frac{\pi}{4 \xi_0}+\frac{\pi}{2}
 \exp(-2\xi_0) \left(1+\frac{1}{2\xi_0}\right)\; .
\end{eqnarray}

With the help of $\mbox{Mapple}^{\textregistered}$ one finds 

\begin{eqnarray}
X_\tau(0) - \frac{2}{ a} &=& 
- \xi_0 +\frac{14}{15}\xi^2_0 -\frac{5}{6}\xi^3_0 +{\cal O}(\xi^4_0)
= -\kappa_0 +\frac{7}{15}\kappa^2_0 a -\frac{5}{24}\kappa^3_0 a^2
+{\cal O}(\kappa^4_0)\; .
\end{eqnarray}

from which we deduce Eq.\ (\ref{D0}).

\item The same method can be used for the expansion of $X_\tau(r)$.
\begin{eqnarray}
\label{Fxi}
X_\tau(r)&=&
\int \frac{d^3\vec{q}}{(2\pi)^3} \tilde{X}_{\tau}(q) 
\exp \left(i\vec{q}\cdot\vec{r} \right)=
 \frac{1}{\pi r } \int_{-\infty}^{+\infty} 
\frac{\sin^2x \sin(2xr/a)}{x(x^2 +\xi_0^2)}\; \frac{1}{1+ \nu(x)}\; dx \; .
\end{eqnarray}

thus, 
\begin{eqnarray}
X_\tau(r)=
 \frac{1}{\pi r } \int_{-\infty}^{+\infty} 
\frac{\sin^2x \sin(2xr/a)}{x(x^2 +\xi_0^2)}\; dx
+ {\cal O}(\xi_0^2) \; ,
\end{eqnarray}

is a piecewise function defined for $r>a$ by the expression
\begin{eqnarray}
\label{lowX}
 X_\tau^<(r)&=&
 \frac{1}{2 r \xi_0^2}\left( 1 - \exp(-\frac{2r}{a} \xi_0)
 -\exp(-2 \xi_0)\sinh(\xi_0 \frac{2r}{a}) \right) + 
 {\cal O}(\xi_0^2)\nonumber \\
&=& \frac{2}{a}-\frac{r}{a^2} -\kappa_0 + {\cal O}(\kappa_0^2)\; ,
\end{eqnarray}
and for $r>a$ by
\begin{eqnarray}
\label{highX}
 X_\tau^>(r)&=&\frac{\sinh^2(\xi_0)}{\xi_0^2}
\frac{\exp(-\frac{2r}{a}\xi_0)}{ r}
= \frac{\sinh^2(\kappa_0 a/2)}{(\kappa_0 a/2)^2}
\frac{\exp(-\kappa_0 r)}{ r} + {\cal O}(\kappa_0^2)\; .
\end{eqnarray}

(\ref{lowX}) and (\ref{highX}) are equivalent to (\ref{psi<}) and (\ref{psi>}).

\item
We give now the expansion of $\omega_1$. We have first

\begin{eqnarray}
\omega_1 \equiv\frac{1}{2} \int \frac{d^d\vec{q}}{(2\pi)^d} \ln 
\left( 1 + \gamma \rho_0\tilde{W}_{\tau}(q) \right)
=  \frac{2}{ \pi^2 a^3} 
 \int_0^\infty  x^2 \ln\left( 1 + \xi_0^2 \frac{\sin^2 x}{x^4} \right) dx
\end{eqnarray}

which can be written as the sum of three terms
\begin{eqnarray}
\omega_1
&=& \frac{2}{ \pi^2 a^3}
 \int_0^\infty  x^2 \left[
\ln\left( 1 + \frac{\xi_0^2 }{x^2} \right) -\frac{\xi_0^2}{x^2 + \xi_0^2}
\right] dx
+ \frac{\xi_0^2}{ \pi^2 a^3} \int_{-\infty}^\infty 
\frac{\sin^2 x}{x^2 +\xi_0^2} dx \nonumber \\
&+&   \frac{1}{ \pi^2 a^3}
\sum_{n=2}^\infty  \frac{(-)^{n-1}}{n}
 \int_{-\infty}^\infty  x^2 \nu^n(x) dx
\end{eqnarray}

The first integral of the last expression can be easily integrated by parts with 
the result 
\begin{eqnarray}
 \int_0^\infty  x^2 \left[
\ln\left( 1 + \frac{\xi_0^2 }{x^2} \right) -\frac{\xi_0^2}{x^2 + \xi_0^2}
\right] dx = \frac{\pi \xi_0^3}{6}\; .
\end{eqnarray}
Thus,
\begin{eqnarray}
\omega_1= \frac{ \xi_0^3}{3\pi a^3} + \frac{1}{ \pi^2 a^3} {\cal I}_0(\xi_0)
- \frac{1}{ \pi^2 a^3}\int_{-\infty}^\infty  x^2 \nu^2(x) dx 
 + {\cal O}(\xi_0^6)\; .
\end{eqnarray}
Using
\begin{eqnarray}
\int_{-\infty}^\infty  x^2 \nu^2(x) dx &=&
\frac{\pi}{16 \xi_0}\left[-9 +
8 \xi_0 + (12 + 8\xi_0) \exp(-2\xi_0) - (3  + 4\xi_0 ) \exp(-4\xi_0)  \right]
+\frac{\pi \xi_0^3}{2}
\nonumber \\
&-&\frac{\pi \xi_0}{2 }+ \pi \xi_0
 \exp(-2\xi_0) \left(1+\frac{1}{2\xi_0}\right)\; .
\end{eqnarray}
we obtain
\begin{eqnarray}
\omega_1=\frac{1}{\pi a^3}\left(\xi_0^2 -\frac{2\xi_0^3}{3} +\frac{7\xi_0^4}
{15} -\frac{\xi_0^5}{3} + {\cal O}(\xi_0^6) \right)
= \frac{\kappa_0^2}{4\pi a} -\frac{\kappa_0^3}{12\pi}+\frac{7\kappa_0^4 a}{240\pi}
-\frac{\kappa_0^5 a^2}{96\pi}+ {\cal O}(\kappa_0^6)\; ,
\end{eqnarray}
which is the result used in Eq.\ (\ref{logarithme}).
\end{enumerate}
%%%%%%%%%%%%%%%%%%%%%%%%%%%%%%%%%%%%%%%%%%%%%%%%%%%%%%%%%%%%%%%%%%%%%%%%%%%%%%%%
%%%%%%%%%%%%%%%%%%%%%%%%%%%%%%%%%%%%%%%%%%%%%%%%%%%%%%%%%%%%%%%%%%%%%%%%%%%%%%%%
\section{Computation of $\omega_2$}

In this section we give the expansion of $\omega_2$ for low $\kappa_0$.
We have
\begin{eqnarray}
\omega_2 &\equiv&  \frac{\langle {\cal H}\rangle_{X_\tau}}{V} 
- \frac{\langle {\cal H}^2\rangle_{X_\tau}-
\langle {\cal H}\rangle^2_{X_\tau}}{2V} \; ,
\end{eqnarray}
with ${\cal H}[\phi] = U_1[\phi]+U_2[\phi]-U_0[\phi]$.

Each cumulant implies several averages over the field $\phi(\vec{r})$ which can all be
deduced from the general average $X \equiv \langle \cos(\lambda_1 \phi_1 ) 
\cos(\lambda_2 \phi_2 ) \cos(\lambda_3 \phi_3 ) \rangle_{X_{\tau}}$ where
$\lambda_1,\lambda_2,\lambda_3$ are real constants.
Starting from
\begin{eqnarray}
X &=& \frac{1}{8} \sum_{\epsilon_1,\epsilon_2,\epsilon_3}
\langle \exp\left( i \left(\lambda_1 \epsilon_1 \phi_1 +
\lambda_2 \epsilon_2 \phi_2 +
\lambda_3 \epsilon_3 \phi_3\right) \right)\rangle_{X_{\tau}}\\
&=& \frac{1}{8} \sum_{\epsilon_1,\epsilon_2,\epsilon_3}
\langle \exp\left( i \int d^3 \vec{r}\;  \phi(\vec{r}) \left(
\epsilon_1 \lambda_1 \delta(\vec{r}-\vec{r}_1) +
\epsilon_2 \lambda_2 \delta(\vec{r}-\vec{r}_2)
+ \epsilon_3 \lambda_3 \delta(\vec{r}-\vec{r}_3) 
\right) \right)
\rangle_{X_{\tau}} \; ,
\end{eqnarray}
where $\epsilon_i=\pm 1 $ and using the fundamental relation for Gaussian integrals
\begin{equation}
\langle \exp\left( i \int d^3 \vec{r}\;  \phi(\vec{r}) \xi(\vec{r}) \right)
 \rangle_{X_{\tau}}=
 \exp(-\frac{1}{2} \int d^3 \vec{r}_1 d^3 \vec{r}_2 \;
 X_{\tau}(\vec{r}_1-\vec{r}_2) \xi(\vec{r}_1) \xi(\vec{r}_2)) \; ,
\end{equation}
one gets

\begin{eqnarray}
X &=& \frac{1}{8} \exp\left(-\frac{X_{\tau}(0)}{2}
\sum_{i=1}^{3}\lambda_i^2 \right) \sum_{\epsilon_i}
\exp \left(
-\lambda_1 \lambda_2 \epsilon_1 \epsilon_2 X_{\tau}(12)
-\lambda_2 \lambda_3 \epsilon_2 \epsilon_3 X_{\tau}(23)
-\lambda_1 \lambda_3 \epsilon_1 \epsilon_3 X_{\tau}(13) \right) \nonumber \\
&=&\exp\left(-\frac{X_{\tau}(0)}{2}
\sum_{i=1}^{3}\lambda_i^2 \right)
\left(\cosh\left(\lambda_1 \lambda_2 X_{\tau}(12) \right) 
     \cosh\left(\lambda_2 \lambda_3 X_{\tau}(23) \right)
      \cosh\left(\lambda_1 \lambda_3 X_{\tau}(13) \right)\right. \nonumber \\
      &-&
\left.      \sinh\left(\lambda_1 \lambda_2 X_{\tau}(12) \right)
      \sinh\left(\lambda_2 \lambda_3 X_{\tau}(23) \right)
      \sinh\left(\lambda_1 \lambda_3 X_{\tau}(13) \right)
\right)      
\end{eqnarray}
from which all useful formula can be obtained.

It is convenient to recast the various contribution of $\omega_2$ in the 
following way
\begin{eqnarray}
\frac{-\langle U_0[\phi]\rangle_{X_\tau}+ \langle U_1[\phi]\rangle_{X_\tau}}{L^3}
&=&-\frac{\rho_0 \Delta_0}{2} 
-\rho_0 \left[\exp(-\Delta_0/2 )- 1\right]
-\frac{\gamma \rho_0}{a} \; ,\\
-\frac{\langle U_0^2  \rangle_{X_\tau} -\langle U_0  \rangle_{X_\tau}^2}{2L^3}&=&
-\frac{\rho_0^2}{4}\int d\vec{r} \psi_\tau^2(r)\; ,\\
\frac{
\langle U_0 U_1  \rangle_{X_\tau} -\langle U_0  \rangle_{X_\tau}\langle U_1  \rangle_{X_\tau}
}{L^3}&=&
\frac{\rho_0^2}{2}\exp(-\Delta_0/2 )
\int d\vec{r} \psi_\tau^2(r)\; ,\\
\frac{\langle U_2[\phi]\rangle_{X_\tau}}{L^3}
-\frac{\langle U_1^2  \rangle_{X_\tau} -\langle U_1  \rangle_{X_\tau}^2}{2L^3}
&=&
-\frac{\rho_0^2}{2}
\left[1-2\exp(-\Delta_0/2)\right]\tilde{h}_0(0)\\
&-&\frac{\rho_0^2}{2}\exp(-\Delta_0)
\int d\vec{r} \left[ g_0(r) \cosh(\psi_\tau(r))
 - 1 \right]\\
&=& -\frac{\rho_0^2}{2}
\left[1-\exp(-\Delta_0/2)\right]^2\tilde{h}_0(0)\\
&-&\frac{\rho_0^2}{2}\exp(-\Delta_0)
\int d\vec{r}  g_0(r) \left[\cosh(\psi_\tau(r))
 - 1 \right] \label{g1}\; .
\end{eqnarray}
where $\Delta_0$ and $\psi_\tau(r)$ are given by Eq.\ (\ref{D0}), (\ref{psi<})
and (\ref{psi>}), and we have introduced $g_0(r)\equiv 1 + h_0(r)$.  

At the lowest order in density the correlation function $g_0^{(0)}(r)=0$ 
for $r<\sigma$ and $1$ otherwise; thus 
\begin{eqnarray}
\label{g2}
\int d\vec{r}  g_0^{(0)}(r) \left[\cosh(\psi_\tau(r))- 1 \right]&=&
\frac{1}{2}\int d\vec{r} g_0^{(0)}(r) \psi_\tau^2(r)
+\sum_{n=2}^\infty 
\int d\vec{r} g_0^{(0)}(r) \frac{\psi_\tau^{2n}(r)}{(2n)!} \nonumber\\
&=&\frac{1}{2}\int d\vec{r} \psi_\tau^2(r)
- \frac{1}{2}\int_{r<\sigma} d\vec{r} \psi_\tau^2(r)
+\sum_{n=2}^\infty 
\int_{r>\sigma} d\vec{r} \frac{\psi_\tau^{2n}(r)}{(2n)!}\; .
\end{eqnarray}                         
 
Gathering together the two precedent results, we obtain Eq.\ (\ref{om2}).
%%%%%%%%%%%%%%%%%%%%%%%%%%%%%%%%%%%%%%%%%%%%%%%%%%%%%%%%%%%%%%%%%%%%%%%%%%%%%%%%
\section{Comparison with Netz-Orland's paper}

In this Appendix we give some technical details allowing the comparison
of Eq.\ (\ref{omz}) with
 the expression of the grand potential in terms of the fugacity calculated
by Netz and Orland.

 Eq.\ (22) of Netz-Orland'paper can be written in our notations ($\lambda \to z,
a \to \sigma$)
\begin{eqnarray}
-\omega_{RPM}=2z-b_{3/2} z^{3/2}\sigma^{3/2}-b_2 z^2 \sigma^3
-b_{5/2}z^{5/2}\sigma^{9/2}-b_{\ln 5/2}z^{5/2}\sigma^{9/2}\ln(z\sigma^3) -\cdots
\end{eqnarray}
where
\begin{eqnarray}
b_{3/2}&=& -\frac{4}{3}\sqrt{2\pi}\epsilon^{3/2}\; ,\\
b_2&=&-2\pi \epsilon^3
-\frac{2\pi}{3}\left(2\epsilon^3
 {\rm Shi}(\epsilon)- \cosh(\epsilon)
[4+2\epsilon^2]- 2 \epsilon 
\sinh(\epsilon)\right)\; \\
b_{5/2}&=&-\frac{(2\pi)^{3/2}}{3}\epsilon^{9/2}
-2\frac{(2\pi \epsilon)^{3/2}}{3}
\left(2\epsilon^3[\Gamma(0,\epsilon)+2\gamma_E+\frac{1}{2}\ln(128\pi\epsilon^3)]
-2\exp(-\epsilon)[2-\epsilon +\epsilon^2]-\frac{59}{12}\epsilon^3\right)\; \label{b52o}\\
b_{\ln 5/2}&=&-2\frac{(2\pi \epsilon)^{3/2}}{3}\epsilon^3
\end{eqnarray}
with $\epsilon \equiv \gamma/\sigma$.

It can be easily checked that $b_{3/2}$ coincides with our coefficient in
 Eq.\ (\ref{omz}). $b_2$ is recovered using the following identity
\begin{eqnarray}
\epsilon^3{\rm Shi}(\epsilon)- \epsilon^2 \cosh(\epsilon)-2\cosh(\epsilon)
 - 2 \epsilon \sinh(\epsilon)= 6{\rm S}(\epsilon)-3\epsilon^2-2
\end{eqnarray}
where ${\rm S}(\epsilon)$ is the series given by Eq.\ (\ref{series}).

$b_{5/2}$ involves the incomplete Gamma function $\Gamma(0,\epsilon)$ which is
related to exponential integral functions by the following relations\cite{Abra}
\begin{eqnarray}
\Gamma(0,\epsilon)&=& {\rm E}_1(\epsilon)\; ,\\
 {\rm E}_{n+1}(\epsilon) &=&\frac{1}{n}\left(
 \exp(-\epsilon) - \epsilon {\rm E}_n(\epsilon) \right)\quad \mbox{for} \quad n>1\; .
\end{eqnarray}

Thus, from $ \epsilon^3 \Gamma(0,\epsilon) - \exp(-\epsilon)\left(\epsilon^2-\epsilon +2 \right)
= -6 {\rm E}_4(\epsilon)  $ we get 
\begin{eqnarray}
\label{b51}
 \epsilon^3 \Gamma(0,\epsilon) - \exp(-\epsilon)\left(\epsilon^2-\epsilon +2 \right)
&=& \epsilon^3 \left( -\ln \epsilon - \gamma_E + \frac{11}{6} \right)
- 2 + 3 \epsilon - 3 \epsilon^2 + 6 \sum_{m=4}^{\infty} \frac{(-\epsilon)^m}{(m-3)m!}\; ,
\end{eqnarray}

where it can be noted that 
\begin{eqnarray}
\label{b52}
 \sum_{m=4}^{\infty} \frac{(-\epsilon)^m}{(m-3)m!} = 
 {\rm S}(\epsilon)- {\rm T}(\epsilon)\; .
\end{eqnarray}

Inserting Eqs\ (\ref{b51}) and (\ref{b52}) in Eq.\ (\ref{b52o}) gives our expression
of the $z^{5/2}$ coefficient in Eq.\ (\ref{omz}).
%%%%%%%%%%%%%%%%%%%%%%%%%%%%%%%%%%%%%%%%%%%%%%%%%%%%%%%%%%%%%%%%%%%%%%%%%%%%%%%%

\end{document}